\documentclass[journal]{IEEEtran}
\usepackage{textgreek}
\usepackage[pdftex]{graphicx}
\usepackage{float}
\graphicspath{{./images}}

\begin{document}
\title{RF Properties and Their Variations in a 3D printed Klystron Circuit and Cavities}

\author{Charlotte Wehner, Julian Merrick, Bradley Shirley, Brandon Weatherford, Garrett Mathesen, and Emilio Nanni
\thanks{This research has been supported by the U.S. Department of Energy (DOE) under Contract No. DE-AC02-76SF00515.}
\thanks{The authors are with the Technology Innovation Directorate, SLAC National Accelerator Laboratory, Menlo Park, CA, 94025 USA}}
\maketitle

\begin{abstract}
Presently, the manufacturing of active RF devices like klystrons is dominated by expensive and time consuming cycles of machining and brazing. In this article we characterize the RF properties of X-band klsytron cavities and an integrated circuit manufactured with a novel additive manufacturing process. Parts are 3D printed in 316L stainless steel with direct metal laser sintering, electroplated in copper, and brazed in one simple braze cycle. Standalone test cavities and integrated circuit cavities were measured throughout the manufacturing process. Un-tuned cavity frequency varies by less than 5\% of intended frequency, and Q factors reach above 1200. A tuning study was performed, and unoptimized tuning pins achieved a tuning range of 138 MHz without compromising Q. Klystron system performance was simulated with as-built cavity parameters and realistic tuning. Together, these results show promise that this process can be used to cheaply and quickly manufacture a new generation of highly integrated high power vacuum devices. 
\end{abstract}


\IEEEpeerreviewmaketitle

\section{Introduction}

\IEEEPARstart{A}{dditive} manufacturing (AM) continues to see growing interest for the manufacturing and development of RF structures and components. Once confined to prototyping, this family of manufacturing techniques is increasingly used to produce devices previously impossible or impractical to manufacture [1]. Many different 3D printing technologies are being explored for RF applications. Common technologies fall into 4 categories.
\begin{figure}
\includegraphics[width=3.5in]{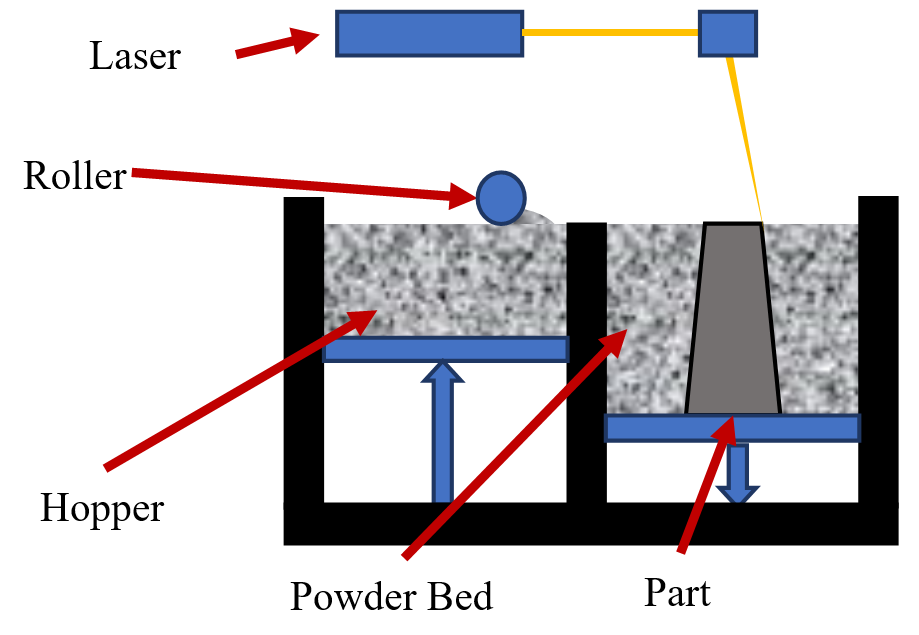}
\caption{Diagram of DMLS printer.}
\end{figure}
\par
Fused Deposition Modeling (FDM), or Fused Filament Fabrication (FFF), involves extruding a thermoplastic through a heated nozzle. The nozzle is precisely moved though the build volume to deposit rapidly solidifying plastic onto the previous layer. While invaluable for rapid prototyping or small scale production due to the very low cost of many FDM 3D printers, the parts this process produces are usually weaker and limited to specific thermoplastics [2]. FDM has been used to manufacture a C-band microwave isolator with a non-printed ferrite bead [3]. However, the limited material selection and poor resolution and accuracy largely prevent FDM from use in the manufacture of RF components. \par

Stereolithography (SLA), sometimes known as vat polymerization, uses an LCD and backlight or laser with a galvanometer to selectively expose a thin layer of photosensitive resin. The part is withdrawn from the vat of photoresist as the next layer is exposed through the bottom of the vat. While also limited to certain plastics, this technology produces strong, accurate, and high-resolution parts while remaining very cost effective [4]. Due to these advantages, significant previous work has demonstrated manufacturing RF devices with SLA. Metal-plated resin waveguides and antennas have been rigorously studied and demonstrate promise for higher frequency devices [5, 6]. 
\begin{figure}[H]
\includegraphics[width=3.5in]{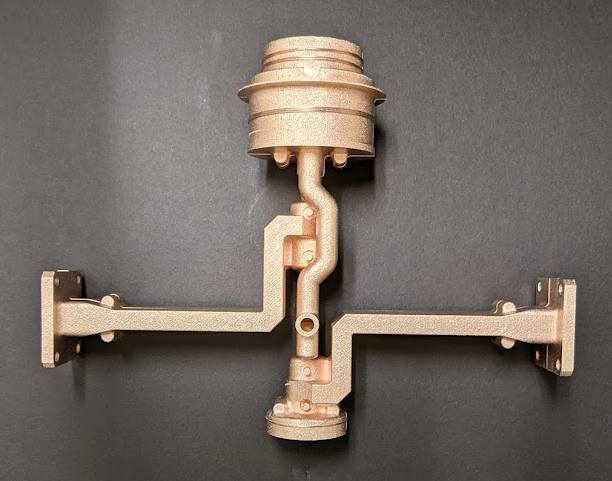}
\caption{Complete klystron circuit with input and output waveguides, pump-out tubes, mounting for an electron gun and collector, and cavity tuning pins.}
\end{figure}
\begin{figure*}
\includegraphics[width=7in]{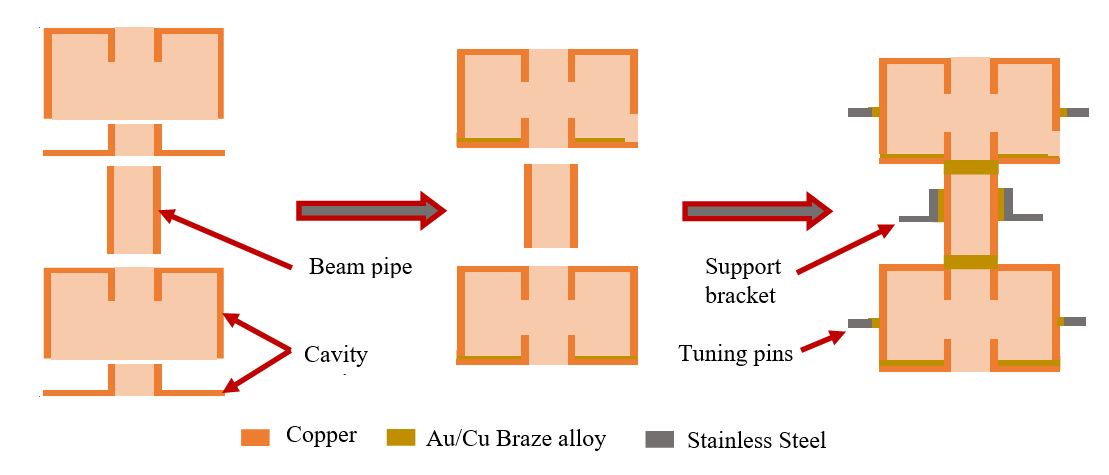}
\caption{Diagram of conventional klystron circuit manufacturing process}
\end{figure*}
\pagebreak
Binder jetting uses a liquid binder to selectively bind together powdered material. Binder is dispensed onto the powder bed by an inkjet to form a 2D layer. Another layer of powder is then deposited. Binder jetting can use a wide variety of powdered materials like metals, ceramics, or polymers. Usually, the binder material is a polymer resin. After printing, the parts often must be post-processed, usually involving baking out the binder and sintering [7]. Binder jetting can produce large, cost effective parts, but so far little work has been done using binder jetting for RF components.

\par
Direct Metal Laser Sintering (DMLS) involves using a high powered laser to selectively sinter metal powder. Many similar technologies like selective laser sintering (SLS) and selective laser melting (SLM) exist. A schematic is presented in Fig. 1. Typically, a laser is directed across the surface of a bed of metal powder. This selectively sinters the powder, forming a layer and adhering it to the previously sintered layer below. Then, a roller deposits another layer of powder onto the bed and the process repeats. Once complete, excess powder must be removed from the part. Often, post-processing techniques like hot isostatic pressing or bead-blasting are performed to improve the mechanical properties and surface quality [8]. Since DMLS can produce metal parts with greater than 99\% density in a variety of materials and alloys [9], it has seen significant use for both prototyping and full-scale production in many industries, including for RF components. For example, waveguides, filters, Tees, and antennas have been demonstrated across many frequencies [10-13].\par

In this article, we present the motivation and process for manufacturing a klystron circuit using DMLS. We also characterize the RF properties of cavities produced with this method. A complete klystron circuit (shown in Fig. 2) and 20 test cavities were manufactured using this process. The RF properties of the test cavities and klystron circuit cavities were measured at multiple stages of the manufacturing process and compared to simulated properties. Also, a tuning study was conducted, and the klystron system performance was simulated with realistic cavity tunings.
\section{Motivation}

A diagram depicting the conventional manufacturing process of a klystron circuit is shown in Fig. 3. Due to the limitations of conventional subtractive manufacturing techniques like CNC Machining, complex structures like reentrant cavities must be split into smaller, simpler parts. These parts must be machined, polished, and cleaned separately before being brazed into subassemblies. These subassemblies may require additional machining and post processing steps. Next, larger subassemblies are brazed together, followed by additional brazing stages to add additional components like supports, tuning pins, and pumpout tubes. Complex assemblies require many stages of machining and brazing, requiring many hours by skilled technicians and effective coordination between manufacturing departments. This labor-intensive process contributes significantly to the high manufacturing cost of such an RF system. Additionally, every braze joint added to the circuit presents an opportunity for vacuum leaks or water leaks from cooling systems. By offloading complexity of manufacturing to automated 3D printing systems, the effective complexity and therefore cost of a system could be significantly reduced. Understanding the RF properties of AM cavities and the properties’ variations is vital to validating this process. 

\section{Manufactured Device and Process}
 \begin{figure*}
\includegraphics[width=7in]{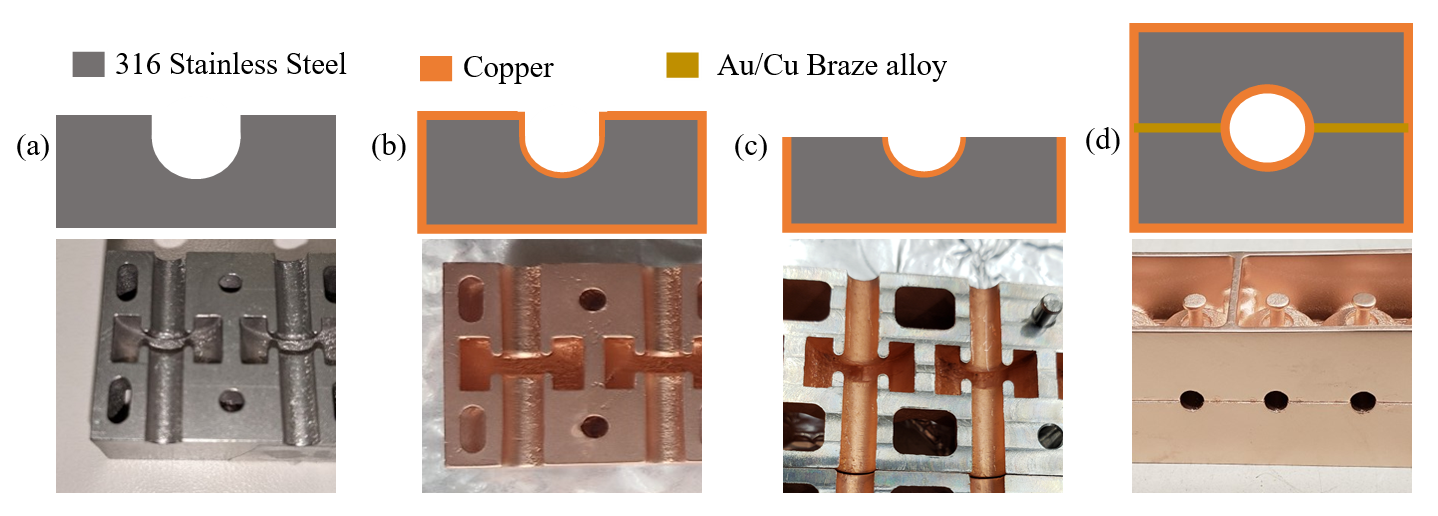}
\caption{Diagram of proposed manufacturing process. (a) Printing, (b) Plating, (c) Face Machining, (d) Brazing}
\end{figure*}
 The proposed manufacturing process leverages the ability of 3D printing to produce extremely detailed, complex parts that would be either completely impossible or prohibitively expensive with conventional methods. The process for a single cavity is depicted in Fig. 4. In DMLS, the cost is driven by the volume of printed material as well as the overall dimensions of the part. Provided that the designer understands the geometric limitations of DMLS, like overhangs and curling, complexity is no longer a limitation. This allows many cavities and other structures to be combined into only two parts. Printing the entire structure as a single part is possible, though impractical due to a reduction of the surface quality of overhanging regions [14] and shadowing effects during electropolishing and plating. \par

Presently, commercially available DMLS materials do not have favorable properties for directly printing RF cavities [15]. To achieve desirable material electrical properties, the parts are printed in 316 stainless steel, then electroplated with copper. The thickness of the copper plating is not crucial, provided that it sufficiently exceeds the skin depth at the target frequency. DMLS exhibits significantly higher surface roughness than conventional milling, so the parts are electropolished before Cu plating. \par
The native surface of the DMLS parts is too rough for brazing with thin, laser-cut shims, so the braze surfaces are first faced flat via conventional milling. Though this is a manual machining process, it consists of only one simple operation per part. Laser-cut 0.002” thick shims of a 25\%/75\% Au/Cu braze alloy are placed between the two flat surfaces of the part. Printed-in holes provide alignment between the two halves. The assembly is held together with stiff wire and brazed in a hydrogen braze furnace. Again, though this is a manual process, it is considerably simpler than the brazing in conventional manufacturing since it only includes two complex parts with well-defined alignment features. Also, reducing the number of braze joints decreases the likelihood of leaks. \par
\begin{figure}[b]
\includegraphics[width=3.5in]{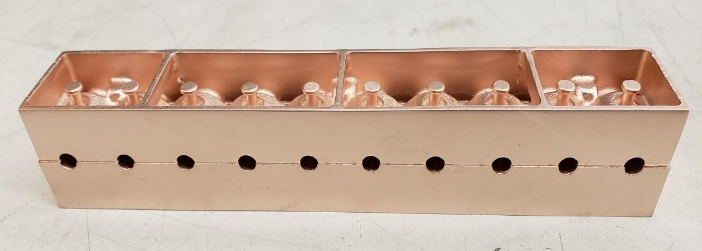}
\caption{Complete brazed test block containing 10 cavities.}
\end{figure}
Two sets of ten test cavities (shown in Fig. 5) were manufactured. These test cavities all had the same design geometry, identical to cavity 2 of the full circuit. In addition to the test cavities, an existing X-band klystron circuit design currently in development at SLAC has been adapted for additive manufacturing. The circuit is designed to output 300 kW at 11.424 GHz. The final device shown in Fig. 6 contains four cavities approximately 20 mm in diameter, the beampipe, input and output waveguides, mounting features for a collector and electron gun, as well as integrated tuning pins. 
\begin{figure}
\includegraphics[width=3.5in]{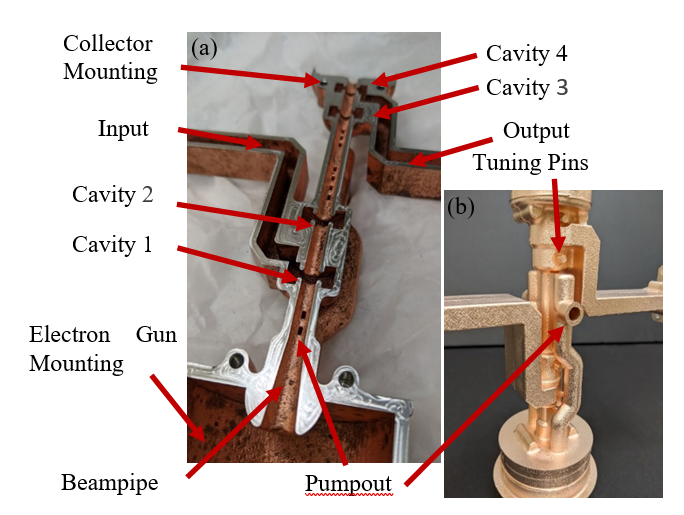}
\caption{Diagram of features included in klystron circuit. (a) Unbrazed half-piece. (b) Full, brazed circuit.}
\end{figure}

\section{Simulation and Characterization}
An eigenmode simulation using ANSYS HFSS was performed to establish the expected cavity performance. The cavity cross-section and E-field is shown in Fig. 7. The expected values for cavity Q and frequency are tabulated in Table 1. Additionally, the entire klystron is simulated using CERN's KlyC code, a 1.5-D large signal simulation code for klystrons to establish expected performance of the system [16]. \par
\begin{figure}
\includegraphics[width=3.5in]{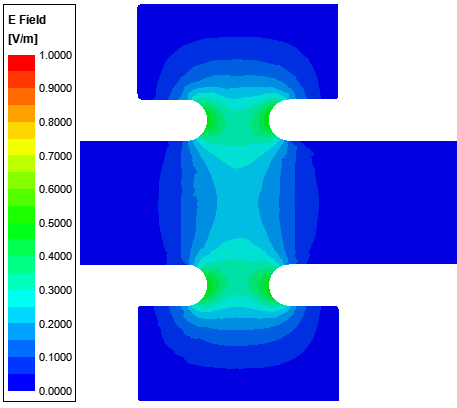}
\caption{Normalized E-field of cavity 3 simulated in HFSS}
\end{figure}
\begin{table}
\centering
\caption{Cavity Resonance and Design Properties}
\begin{tabular}{lllll}
\hline
\hline
\textbf{Cavity}            & 1      & 2     & 3      & 4     \\ \hline
\textbf{Frequency   (GHz)} & 11.424 & 11.44 & 11.495 & 11.41 \\ \hline
\textbf{Quality Factor}    & 200    & 2000  & 2000   & 125  \\
\hline
\hline
\end{tabular}
\end{table}
Cavities were measured using two E-field probes on an Agilent N5241A network analyzer. Each probe was positioned near the irises of each successive cavity using manual slide stages. The peak resonance of each cavity was determined by inserting both probes completely into each cavity, then slowly removing them while monitoring the reflection and transmission coefficient curves. Once it was obvious that removing each probe further did not affect the resonant frequency (i.e., the cavities are not being detuned) a measurement was taken. Since the metal body of the probes extended through the entire beampipe, except in the cavity under test, other cavities are shorted. A waveguide load was placed on both the input and output waveguides during testing. \par 
Every cavity of the full klystron circuit was measured both before and after brazing. The Q factor and frequency was calculated from the peak in the S12 transmission coefficient. To characterize the variation inherent in every manufacturing step, the 20 test cavities were characterized immediately after printing, after electropolishing, after plating, after face machining, and after brazing. 

\section{Test Cavity Results}
The as-printed resonance for all 20 test cavities is shown in Fig. 8 and Fig. 9. Note that the simulated frequency and Q accounts for the additional material which is to be removed during the face-machining step. There is a significant spread in both frequency and Q, however, observe that Q factor is uniformly lower, and frequency uniformly higher than the design intent. These trends, once characterized, would allow the designer to compensate by designing the cavities to have a higher Q factor and lower frequency than intended. \par
\begin{figure}
\includegraphics[width=3.5in]{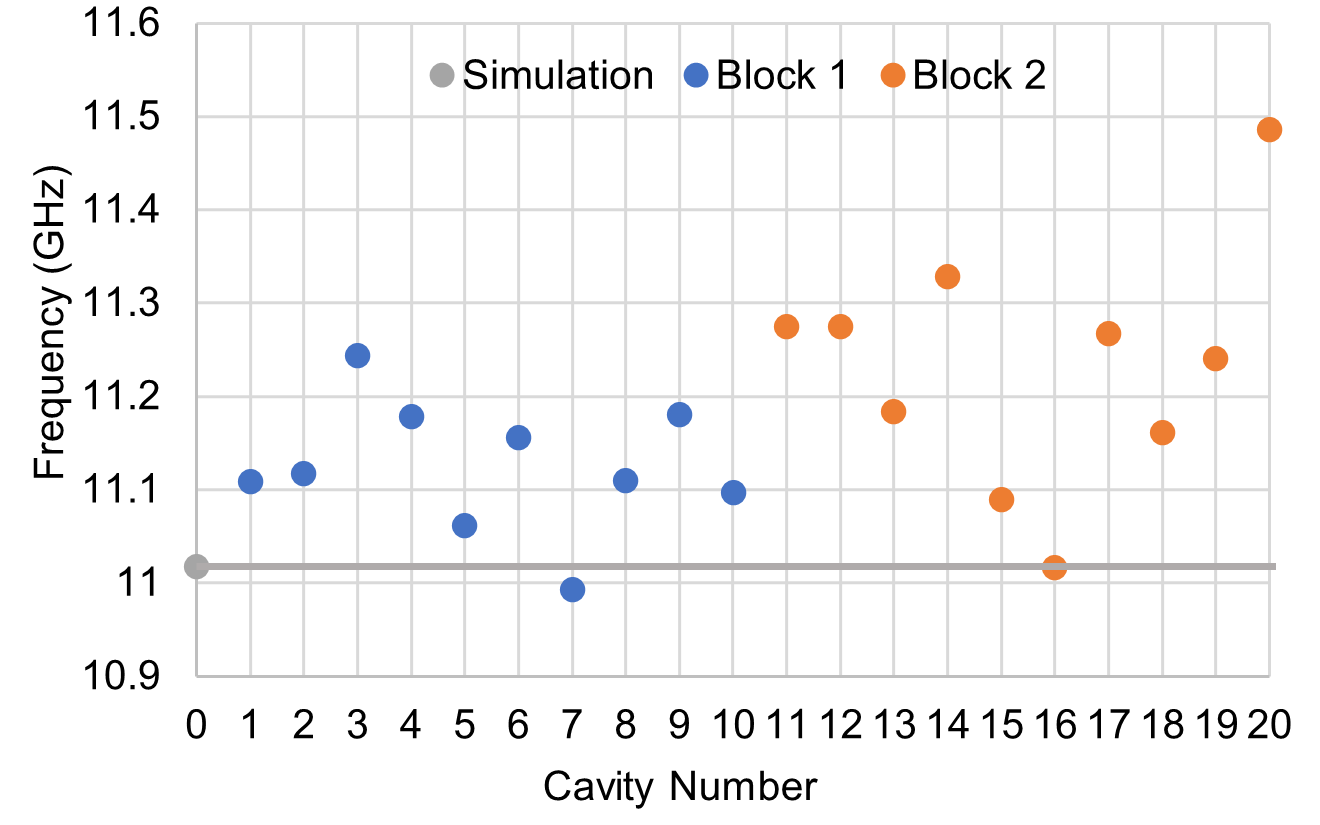}
\caption{Frequency of all test cavities measured after printing. Simulated frequency is shown as horizontal grey line.}
\end{figure}
\begin{figure}
\includegraphics[width=3.5in]{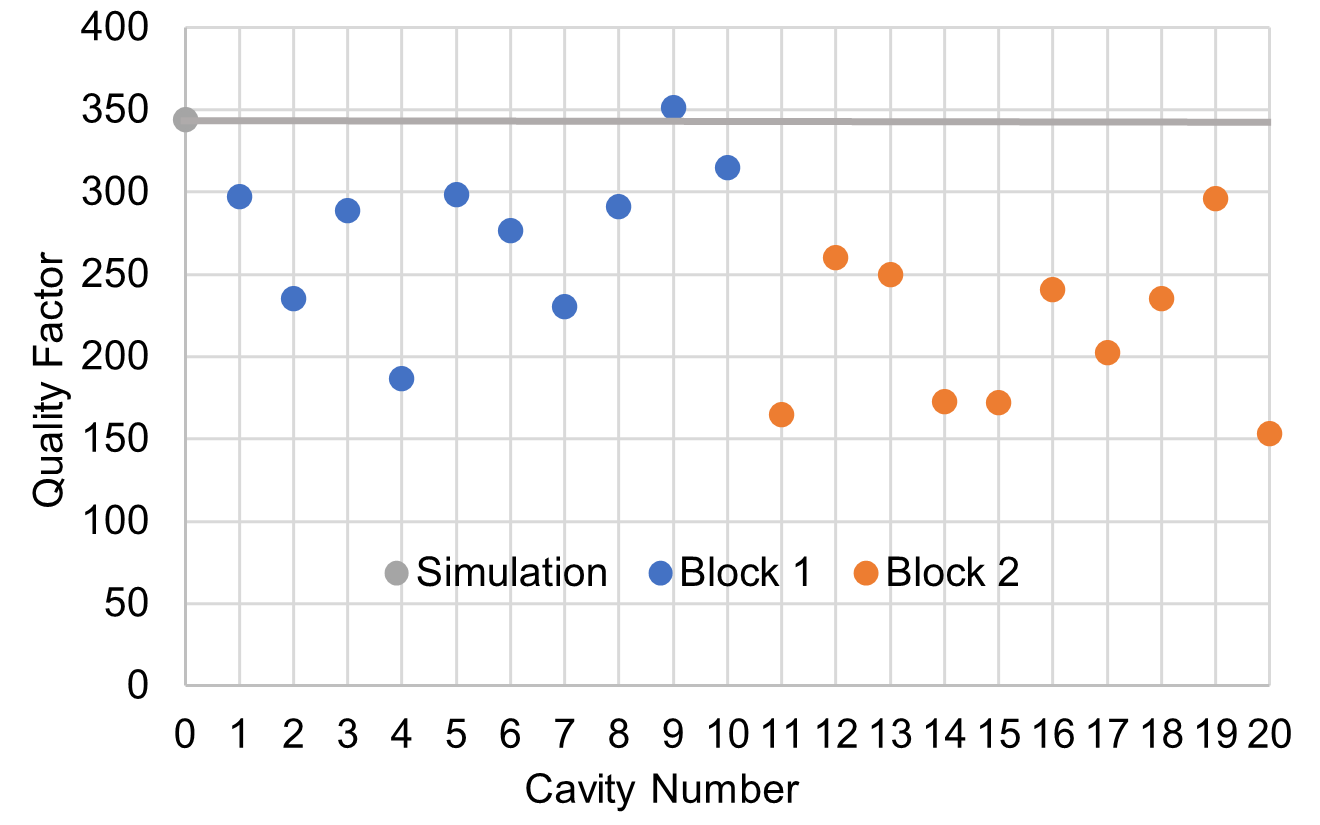}
\caption{Q factor of all test cavities measured after printing. Simulated Q factor is shown as horizontal grey line.}
\end{figure}
\begin{table}[]
\centering
\caption{Test Cavity Properties Throughout Manufacturing Process}
\begin{tabular}{p{40pt}lllp{40pt}}
\hline
\hline
 & \textbf{As Printed} & \textbf{Polished} & \textbf{Plated} & \textbf{Face Machined*} \\ \hline
\textbf{Mean Frequency (GHz)} & 11.178              & 11.378            & 11.380          & 11.565                  \\ \hline
\textbf{Frequency \textsigma{} (GHz)}  & 0.115               & 0.181             & 0.125           & 0.107                   \\ \hline
\textbf{Mean Q Factor}        & 246                 & 313               & 1961            & 1317                    \\ \hline
\textbf{Q Factor CV}          & 0.228               & 0.265             & 0.243           & 0.48                    \\ \hline
\hline
\end{tabular}
\vspace{3pt}
\par
* Data only from 10 test cavities. Other data from all 20.
\end{table}
\begin{figure}[t]
\includegraphics[width=3.5in]{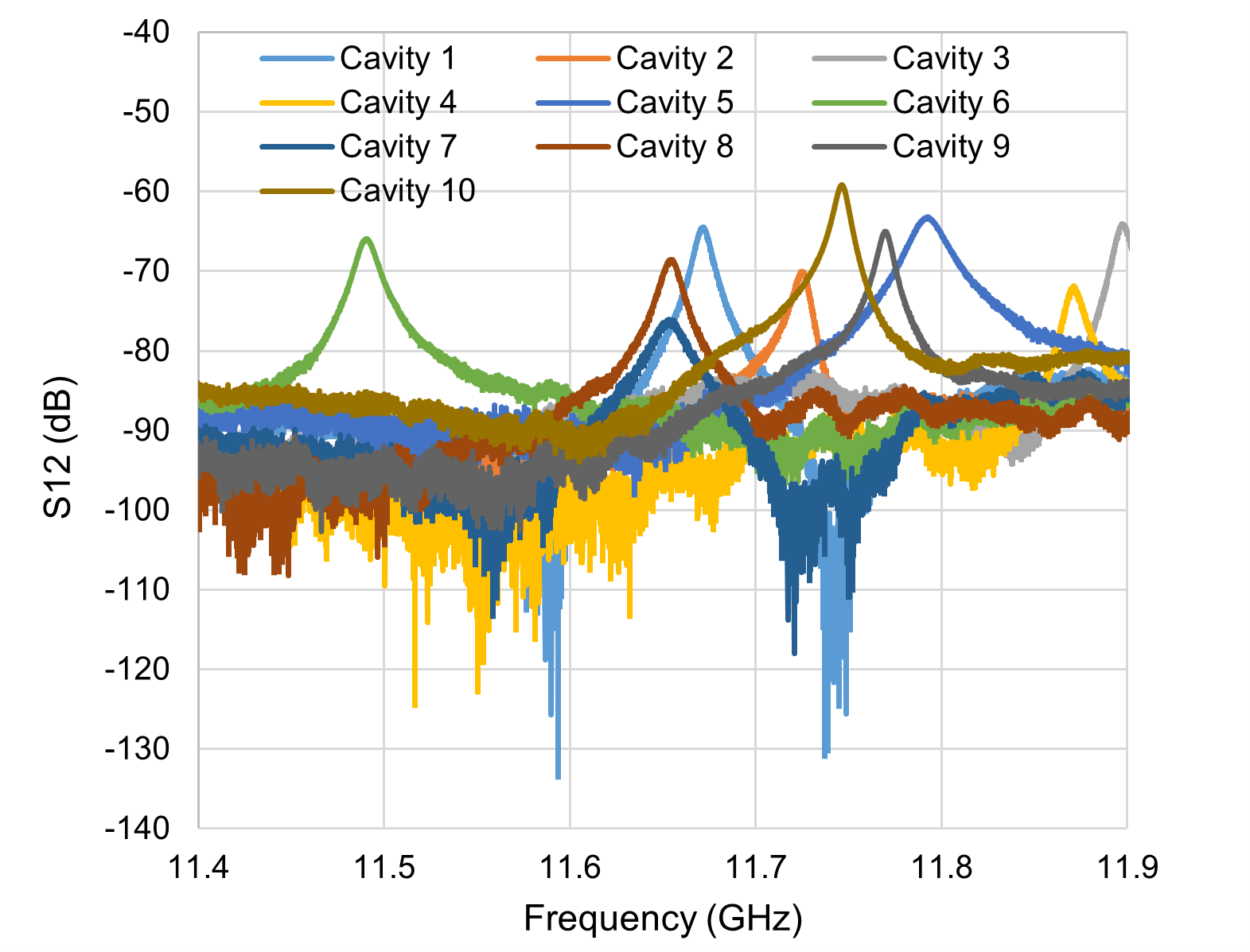}
\caption{Transmission coefficient of 10 test cavities measured after brazing.}
\end{figure}
The cavity parameters before and after each major manufacturing step are shown in Table 2. After polishing, the frequency and frequency standard deviation rise. The Q also rises, but this is expected given that the lowered surface roughness would contribute to a lower effective surface resistance. After plating, Q rises significantly since the cavity surface is copper. Standard deviation for both frequency and Q decreases. After machining, the frequency rises significantly as the cavity dimensions are brought to their final values. The Q factor decreases, and the Q factor standard deviation rises significantly, though this is not a direct comparison as the face machining data is only from ten out of twenty test cavities.\par
The final brazed cavity resonance of ten cavities is shown in Fig. 10. There is significant variation in both frequency and Q factor. However, most of the cavities have a frequency within 150 MHz of each other. During the printing process, the 3D model provided to the printer must be sliced into layers, then converted into a series of instructions for the printer to execute to create the part. Combined with phenomena inherent to the physical process of sintering powder, these factors mean that the actual printed part geometry is not identical to the model provided to the printer. Two fundamental types of variations are identified. \par
Firstly, there are consistent variations which similarly affect all parts. For example, it has been shown that small holes in DMLS parts are consistently printed smaller than the model [14]. These variation work to offset the mean frequency and mean quality factor from the theoretical values. Once these variations are characterized, the designer can compensate. For example, the mean frequency of the final cavities as shown in Table 2 is 141 MHz higher than the theoretical frequency. To compensate, the designer could target a frequency of 11.283 GHz. Assuming the offset remains consistent, the actual mean frequency of those cavities would be much closer to desired frequency of 11.424 GHz.\par
Secondly, inconsistent variations which may differently affect each theoretically identical part. These inconsistent errors contribute to the imprecision of the part and work to increase the standard deviation observed in the cavities’ frequency and Q factor. Since these variations are inherently not predictable, they provide a more difficult challenge to creating accurate cavities. Work must go into both limiting these variations and compensating for them via tuning. \par

\section{Tuning Study}
Tuning conventionally-manufactured copper cavities via mechanical deformation is relatively well-understood. However it is expected that the different mechanical properties of stainless steel, as well as the single-piece design of print-in-place tuning pins, will behave differently during tuning. To quantify this behavior, a preliminary tuning study was performed by impacting one of the completed, brazed test cavities multiple times. Between each impact, the cavity resonance was measured using the standard technique. The results of this study are shown in Fig. 11. Through 5 rounds of tuning, the cavity frequency was raised more than 150 MHz. These traces also indicate that the Q factor was compromised.
\begin{figure}[t]
\includegraphics[width=3.5in]{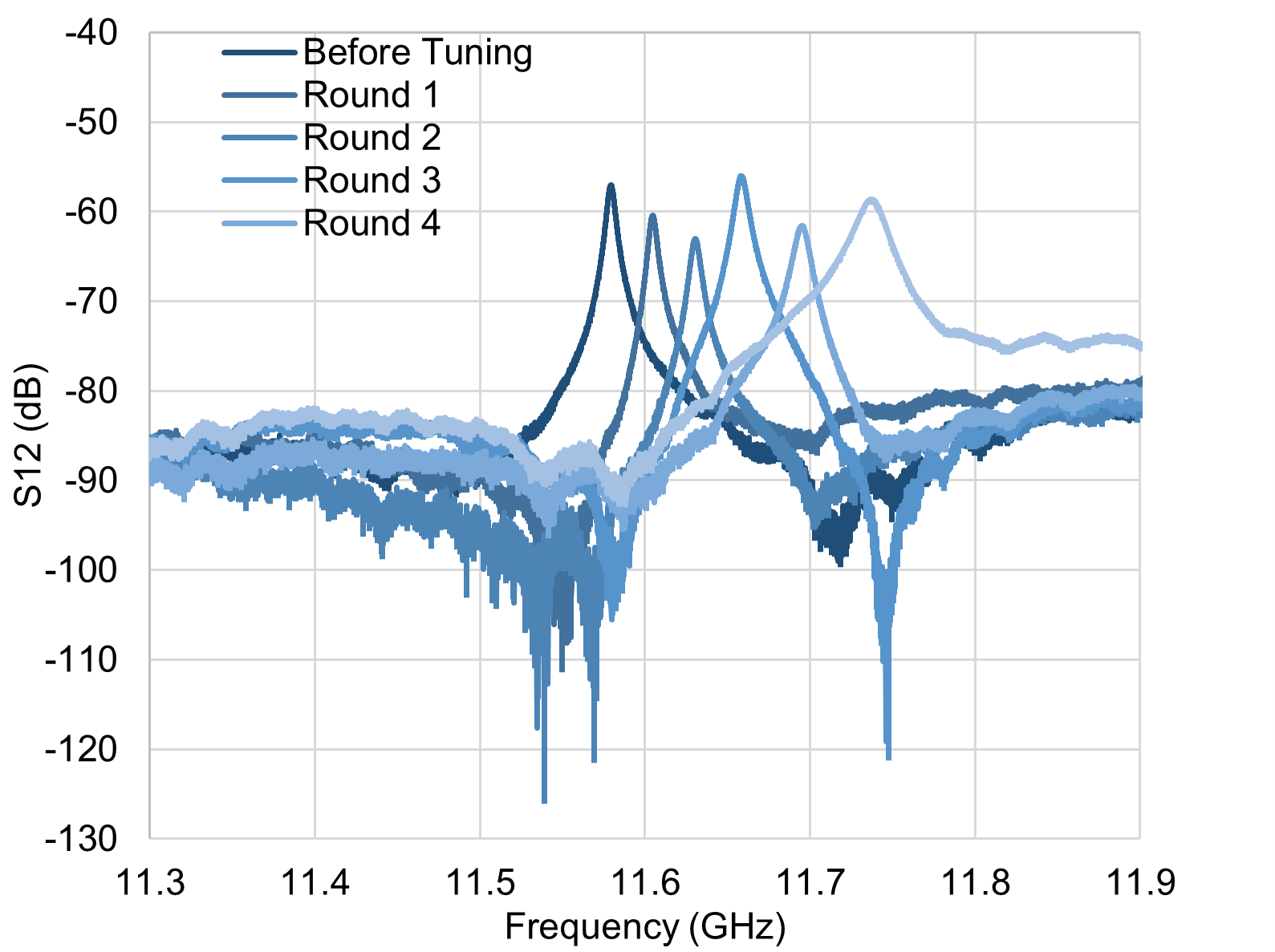}
\caption{Transmission coefficient of test cavity measured after successive rounds of tuning.}
\end{figure}
Additionally, as can be seen in Fig. 12, the tuning pin is significantly deformed. In a different tuning test, the pin broke completely through the cavity wall, as shown in Fig. 13.\par
\begin{figure}
\includegraphics[width=3.5in]{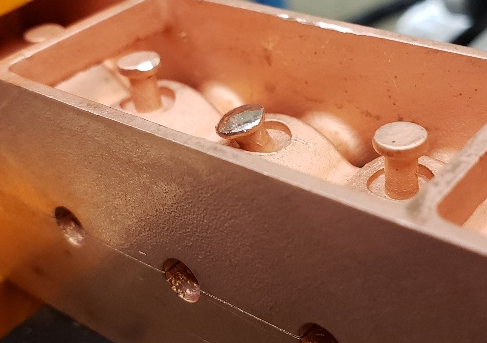}
\caption{Deformed tuning pin after tuning test}
\end{figure}
\begin{figure}
\includegraphics[width=3.5in]{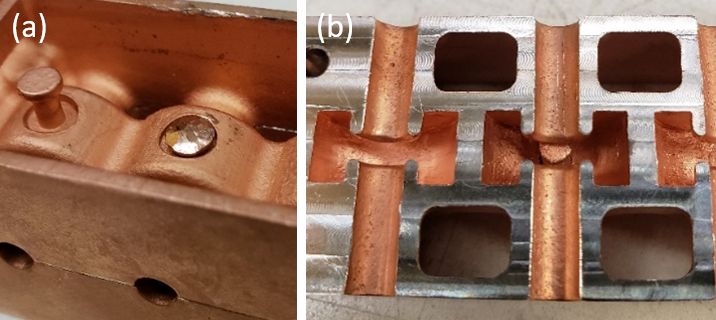}
\caption{Damaged cavity after failed tuning test. (a) Tuning pin head on exterior of cavity. Undamaged pin is shown on the left. (b) Interior of damaged cavity with pin protruding. Undamaged cavity is shown on the left.}
\end{figure}
\begin{table*}[]
\caption{Test Cavity Tuning Results}
\centering
\begin{tabular}{lllllllll}
\hline
\hline
\multicolumn{1}{l|}{}                & \multicolumn{3}{l|}{\textbf{Before}}                     & \multicolumn{2}{l|}{\textbf{After}}         & \multicolumn{2}{l}{}                       &                  \\ \hline
\multicolumn{1}{l|}{\textbf{Cavity}} & Frequency (GHz)          & \multicolumn{2}{l|}{Q}        & Frequency (GHz) & \multicolumn{1}{l|}{Q}    & \multicolumn{2}{l}{Frequency Change (MHz)} & Deformation (\textmu{}m) \\ \hline
\multicolumn{1}{l|}{\textbf{1}}      & 11.671                   & \multicolumn{2}{l|}{1284}     & 11.778          & \multicolumn{1}{l|}{1220} & \multicolumn{2}{l}{107}                    & 516              \\ \hline
\multicolumn{1}{l|}{\textbf{2}}      & 11.897                   & \multicolumn{2}{l|}{1250}     & 11.84           & \multicolumn{1}{l|}{1275} & \multicolumn{2}{l}{-57}                    & -347             \\ \hline
\multicolumn{1}{l|}{\textbf{3}}      & 11.725                   & \multicolumn{2}{l|}{1407}     & 11.819          & \multicolumn{1}{l|}{1195} & \multicolumn{2}{l}{94}                     & 511              \\ \hline
\multicolumn{1}{l|}{\textbf{4}}               & 11.871                   & \multicolumn{2}{l|}{1180}     & 11.988          & \multicolumn{1}{l|}{1095} & \multicolumn{2}{l}{117}                    & 438              \\ \hline
\multicolumn{1}{l|}{\textbf{5}}               & 11.608                   & \multicolumn{2}{l|}{1195}     & 11.575          & \multicolumn{1}{l|}{1197} & \multicolumn{2}{l}{-33}                    & -297             \\ \hline
\multicolumn{1}{l|}{\textbf{6}}               & 11.49                    & \multicolumn{2}{l|}{967}      & 11.457          & \multicolumn{1}{l|}{962}  & \multicolumn{2}{l}{-33}                    & -320             \\ \hline
\multicolumn{1}{l|}{\textbf{7}}               & 11.583                   & \multicolumn{2}{l|}{566}      & 12.029          & \multicolumn{1}{l|}{1101} & \multicolumn{2}{l}{446*}                   & 648              \\ \hline
\multicolumn{1}{l|}{\textbf{8}}               & 11.655                   & \multicolumn{2}{l|}{906}      & 11.619          & \multicolumn{1}{l|}{1030} & \multicolumn{2}{l}{-36}                    & -350             \\ \hline
                                     & \multicolumn{2}{l}{Average increase (MHz):} & \multicolumn{2}{l}{106}      & \multicolumn{2}{l}{Average Decrease (MHz):}      & \multicolumn{2}{l}{-32}                \\ \hline \hline
\end{tabular}
\par
\vspace{3pt}
* Excluded from average due to questionably large shift
\end{table*}
An additional tuning study was conducted on 8 of the test cavities. The results are presented in Table 3. For this study, a tuning tool was fabricated. The tool grasps the head of each tuning pin and can smoothly push or pull the pin with a screw. Four cavities each were tuned upwards and downwards to the maximum extent permitted by the tuning tool. The resonance of each cavity was measured before and after tuning. After tuning, the cavities were cut open and the cavity wall acted on by the tuning pin was measured on a laser confocal microscope. The deformation in the cavity wall caused by tuning was measured. These results establish an expected tuning bandwidth of 138 MHz with current tuning structures. As expected, the distance the cavity wall is deformed correlates to the frequency shift. Further, Q factor does not appear to be compromised through tuning, with some cavities even increasing in quality factor after tuning. Together, these results suggest that even with current tuning pin geometry, any cavity frequency spread less than 138 MHz will allow cavities to be tuned to an exact frequency. Note that these tuning structures were completely unoptimized, and by affecting a larger portion of the cavity wall and therefore displacing more cavity volume, it is likely that significantly more effective tuning structures can be designed. 

\section{Klystron Circuit Results}
The klystron circuit underwent a helium leak check which showed a good braze joint with no leaks. Further, a DMLS sample was placed in vacuum to assess the material’s suitability under vacuum. No significant outgassing was observed. \par
\begin{figure}
\includegraphics[width=3.5in, height=2.5in]{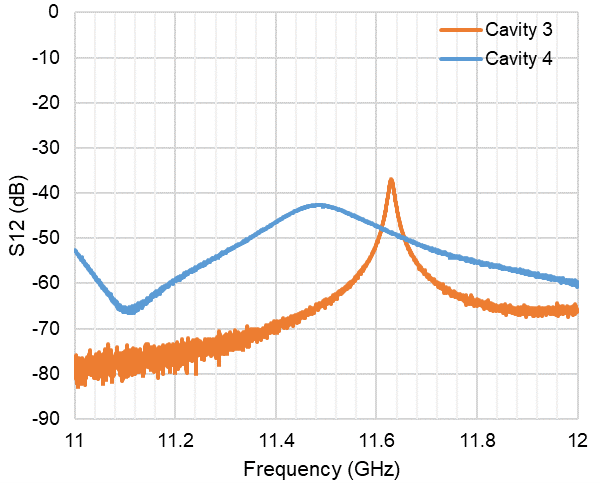}
\caption{Measured transmission coefficients of cavities 3 and 4 measured after brazing.}
\end{figure}
The measured resonant peaks for circuit cavities 3 and 4 are shown in Fig. 14. Both cavities show significant resonance. Cavity 3 is very clear and shows relatively high Q. Cavity 4 is less clear, likely due to coupling into the connected output waveguide. Determining the peak on cavity 4 was difficult due to the proximity to the end of the beampipe. The resonance of all circuit cavities both before and after brazing are shown in Fig. 15 and 16. There is significant variation both between the cavities and between each cavity and its intended frequency. Cavity 2 exhibits the largest difference between intended and actual frequency at 4.6\%.\par 
\begin{figure}
\includegraphics[width=3.5in]{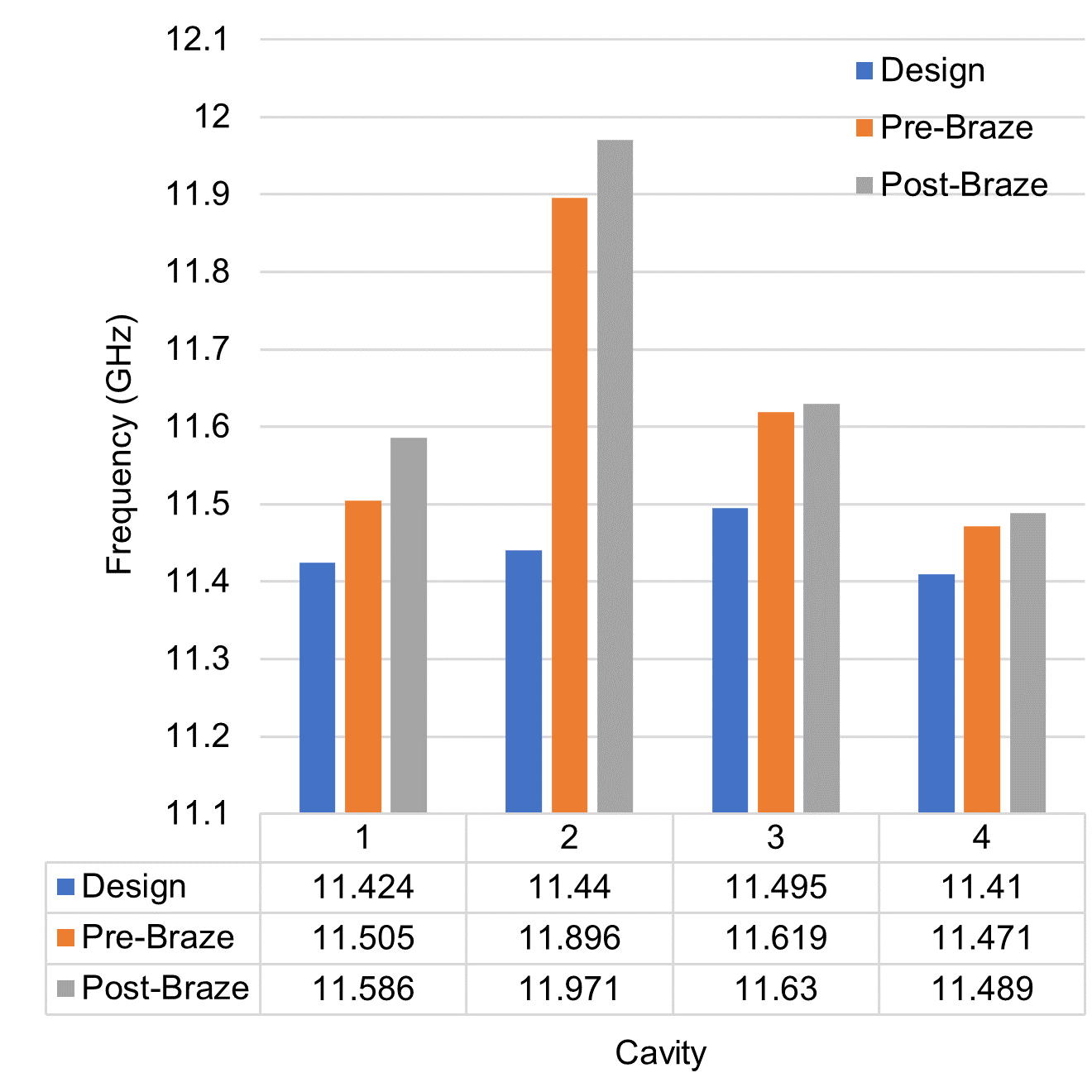}
\caption{Resonant frequency of klystron circuit cavities as simulated, before brazing, and after brazing. }
\end{figure}\begin{figure}
\includegraphics[width=3.5in]{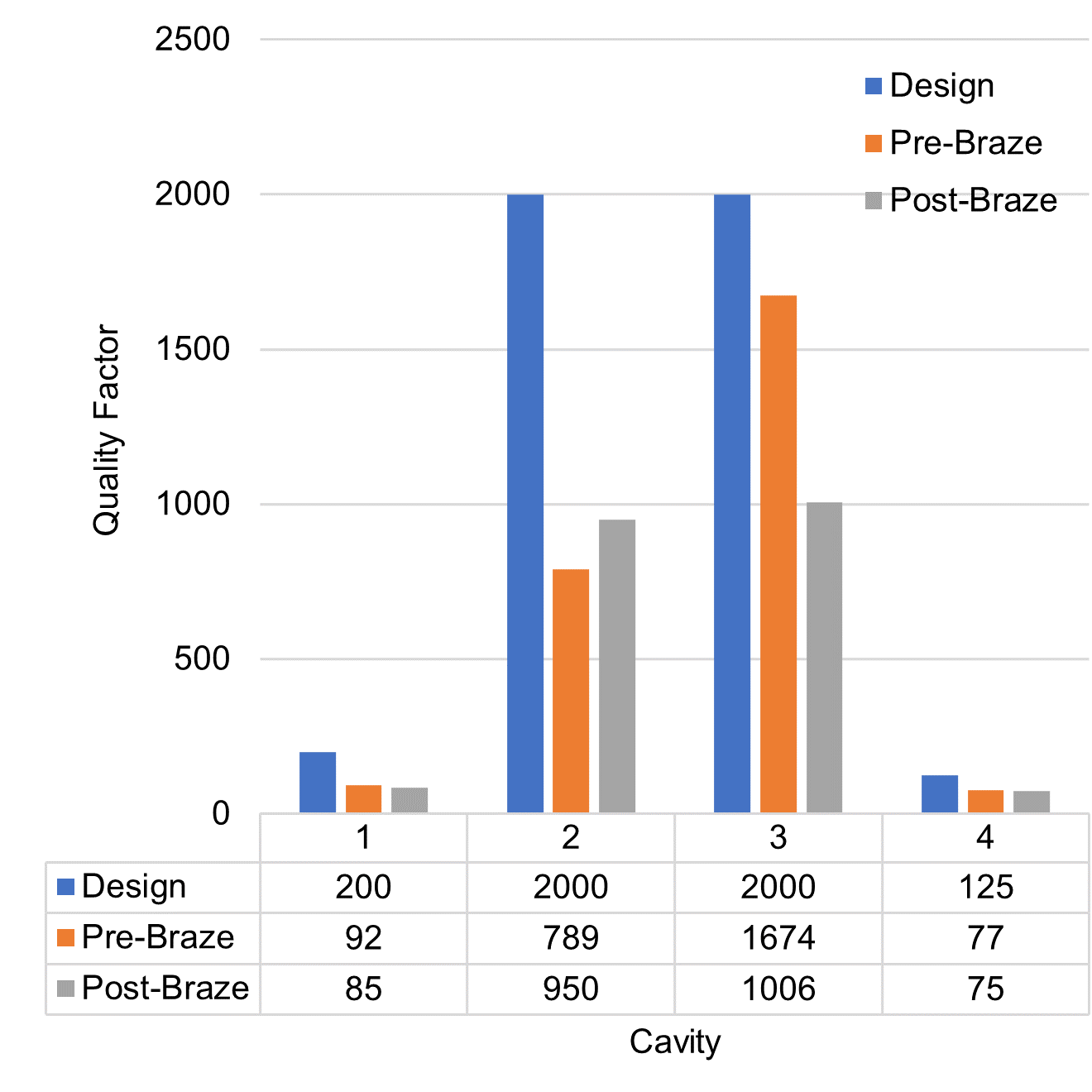}
\caption{Quality factor of klystron circuit cavities as simulated, before brazing, and after brazing. }
\end{figure}
The measured Q factor shows significant promise. First, the final post-braze resonance is very nearly half of the design Q for every cavity. Second, with a Q factor of around 1000, cavities 2 and 3 demonstrate that this manufacturing process can indeed create cavities with Q factors on the order necessary for klystrons. \par
To inform further testing and determine expected performance, the klystron was simulated, and an optimization was performed. Two potential tuning solutions are presented. In both cases, the operational frequency of the klystron has been changed from 11.424 GHz. Allowing flexibility in this parameter enabled considerably higher output power for this test device. In a production klystron, the designer could target a specific operational frequency by offsetting the designed frequency of the cavities. In these simulations, it is assumed that tuning does not affect the Q factor.\par
\begin{table}[h]
\caption{Klystron Tuning Solution 1}
\centering
\begin{tabular}{lll}
\hline
\hline
\textbf{Cavity}     & \textbf{Frequency (GHz)} & \textbf{Tuning Needed   (MHz)} \\ \hline
\textbf{1} & 11.955                   & 369                            \\ \hline
\textbf{2} & 11.971                   & 0                              \\ \hline
\textbf{3} & 12.000                   & 370                            \\ \hline
\textbf{4} & 11.941                   & 542                            \\ \hline
\hline
\end{tabular}
\end{table}
Solution 1 is shown in Table 4. Due to the large frequency offset of cavity 2, the rest of the cavities must be tuned by nearly 400 MHz, outside the established tuning range. This would require refined tuning structures which permit a larger tuning range. However, this solution allows an output power of 257 kW at a frequency of 11.955 GHz. \par
\begin{table}[h]
\caption{Klystron Tuning Solution 2}
\centering
\begin{tabular}{lll}
\hline
\hline
\textbf{Cavity}     & \textbf{Frequency (GHz)} & \textbf{Tuning Needed   (MHz)} \\ \hline
\textbf{1} & 11.620                   & 34                            \\ \hline
\textbf{2} & 11.971                   & 0                              \\ \hline
\textbf{3} & 11.630                   & 0                            \\ \hline
\textbf{4} & 11.620                  & 131                            \\ \hline
\hline
\end{tabular}
\end{table}
Solution 2 is shown in Table 5. In this solution, cavity 2 is untuned, so the large frequency offset means the cavity does not significantly participate in the klystron operation. Because of this, the remaining cavities need to be tuned significantly less, even less than the tuning range achievable with current tuning structures. However, effectively removing cavity 2 reduces the power output to 55 kW at 11.620 GHz.

\section{Discussion}
Although early in development, the measured RF properties and klystron simulations show promise that the manufactured klystron circuit can operate as a viable RF source. Additionally, the properties of the RF cavities indicate that only incremental progress is necessary to create consistent and accurate AM cavities as part of a tightly integrated and cost-effective device. However, further testing is necessary to prove vacuum tubes manufactured with this process can be functional. \par
The manufactured device should be pumped down to determine whether the structure can achieve a low enough base pressure. The slightly porous structure of the DMLS stainless steel as well as brazing in a hydrogen atmosphere may introduce additional outgassing and adsorption.\par 
Due to the observed inconsistencies, further investigation and reduction of the large variations present in cavity frequency and quality is critical to the successes of this manufacturing process. These inconsistencies likely arise through multiple phenomena. \par
Firstly, the surface roughness introduces variation into the surface resistance of cavity walls, as well as small-scale geometry which could impact RF behavior. Multiple routes for further development exist. DMLS was carried out by a commercial vendor and was likely optimized for acceptable and repeatable performance across many different metrics relevant to many different industries. Refining DMLS process parameters specifically for this application could yield lower-roughness parts more suitable for RF structures. Additionally, post-processing steps like electropolishing or bead blasting could be investigated in more detail to reduce roughness after printing. \par
Second, DMLS has been shown to introduce bulk geometry errors, such as small holes being smaller than expected. Further study should carefully measure the as-printed geometry and characterize these errors. It would be especially valuable to know if and how the designer could compensate for these errors before printing. \par
With only incremental improvement in the variation of RF properties and tuning bandwidth, X-band AM cavities can be cost-effectively produced to target a specific frequency and Q factor. AM provides the flexibility to quickly achieve this incremental improvement and integrate the resulting cavities into the next generation of klystrons and other high power X-band devices.

\end{document}